\documentclass{iaus}
\usepackage{graphicx}

  \checkfont{eurm10}
  \iffontfound
    \IfFileExists{upmath.sty}
      {\typeout{^^JFound AMS Euler Roman fonts on the system,
                   using the 'upmath' package.^^J}%
       \usepackage{upmath}}
      {\typeout{^^JFound AMS Euler Roman fonts on the system, but you
                   dont seem to have the}%
       \typeout{'upmath' package installed. iaus.cls can take advantage
                 of these fonts,^^Jif you use 'upmath' package.^^J}%
      }
  \else
  \fi

  \checkfont{msam10}
  \iffontfound
    \IfFileExists{amssymb.sty}
      {\typeout{^^JFound AMS Symbol fonts on the system, using the
                'amssymb' package.^^J}%
       \usepackage{amssymb}%

      }{}
  \fi

  \IfFileExists{amsbsy.sty}
    {\typeout{^^JFound the 'amsbsy' package on the system, using it.^^J}%
     \usepackage{amsbsy}}
    {}

\newsavebox{\astrutbox}
\sbox{\astrutbox}{\rule[-5pt]{0pt}{20pt}}

\title[Black Hole Masses from Reverberation Measurements]
{Black Hole Masses from Reverberation Measurements}

\author[B.M.\ Peterson]%
{Bradley M. Peterson}

\affiliation{$^1$Department of Astronomy, The Ohio State
University, 140 West 18th Avenue, Columbus, OH 43210, USA
email: peterson@astronomy.ohio-state.edu}

\pubyear{2004}
\volume{222}
\pagerange{1--8}
\date{?? and in revised form ??}
\jname{The Interplay among Black Holes, Stars and ISM \\in Galactic Nuclei}
\editors{Th. Storchi Bergmann, L.C. Ho \& H.R. Schmitt, eds.}
\begin{document}

\maketitle

\begin{abstract}
We have reanalyzed in a consistent way 
existing reverberation data for 35 AGNs
for the purpose of refining the black hole masses
derived from these data. We find that the precision (or random component of
the error) of reverberation-based black hole mass measurements
is typically around 30\%, comparable to the
precision attained in measurement of black hole masses in
quiescent galaxies by gas or stellar dynamical methods.
As discussed in this volume by Onken et al., 
we have established an absolute calibration for
AGN reverberation-based masses by assuming that AGNs and
quiescent galaxies follow an identical relationship between
black hole mass and host-galaxy bulge velocity dispersion.
The scatter around this relationship implies that the
typical systematic uncertainties in reverberation-based black hole
masses are smaller than a factor of three.
We present a preliminary version of a mass--luminosity relationship
that is much better defined than any previous attempt.
Scatter about the mass--luminosity
relationship for these AGNs appears to be real and
could be correlated with either Eddington ratio or source inclination.
\end{abstract}

\firstsection 
\section{Introduction}
While black holes have been invoked since the early days of
quasar research as the principal ingredient in AGNs,
it is only within the last several years that it has become
possible to measure the masses of the central objects
in galactic nuclei. At the present time, there has not yet been a
definitive detection of the relativistic effects that 
would be required for unambiguous identification of a singularity,
although studies of the Fe K$\alpha$ emission line in the
X-ray spectra of AGNs currently afford some promise
(e.g., Reynolds \& Nowak 2002). However, it seems to be true 
that the centers of both active and quiescent galaxies host 
supermassive objects that must 
be so compact that other alternatives are very unlikely.  

Black hole masses are measured in a number of 
ways. In Type 1 active galaxies, reverberation 
mapping (hereafter RM; Blandford \& McKee 1982; Peterson 1993) of the 
broad-line region (BLR) can be used to determine the central 
masses. RM is the only method of 
black hole mass measurement that does 
not depend on high angular resolution, so it is of special interest 
as it is extendable in principle to both very high and very 
low luminosities and to objects at great distances. Moreover, 
RM studies reveal the existence of simple scaling 
relationships that can be used to anchor secondary methods of 
mass measurement, thus making it possible to provide estimates 
of the masses of large samples of quasars, including
even very distant quasars, based on  relatively simple 
spectral measurements (e.g., Vestergaard 2002, 2004; McLure \& Jarvis 2002).

The evidence that RM-based black hole masses are
valid is twofold:
\begin{enumerate}
\item There is an anticorrelation between emission-line time
lag $\tau$, which measures the size of the line-emitting region by light-travel
time, and emission-line width $\Delta V$ that is
consistent with the virial prediction $\tau \propto \Delta V^{-2}$.
Relative to lower ionization lines, 
higher ionization lines are both broader
and closer to the central source, consistent with ionization
stratification of the BLR and dynamics that are dominated by
a central mass (Peterson \& Wandel 1999, 2000; Onken \& Peterson 2002;
Kollatschny 2003).
\item There is a relationship (Gebhardt et al. 2000b;
Ferrarese et al.\ 2001) between the 
RM-based black hole masses and the host
galaxy stellar velocity dispersion $\sigma_*$ similar to
that seen in quiescent galaxies (Ferrarese \& Merritt 2000;
Gebhardt et al.\ 2000a).
\end{enumerate}

A difficulty with the existing RM database is the
lack of uniformity in how the measurements have been made.
This leads to considerable
scatter in many of the relationships among various
AGN properties. We have therefore undertaken a complete
reanalysis of existing RM data with two
goals in mind:
\begin{enumerate}
\item In order to improve the {\em precision}
(i.e., reduce the random errors) of AGN black hole 
mass measurements, we reanalyzed all of the readily 
available RM data to determine the best measures of 
time lag and line width for these studies. 
\item In order to improve the {\em accuracy}
(i.e., reduce the systematic errors) of AGN black hole 
mass measurements, we obtained high-precision 
measurements of $\sigma_*$ to determine the
$M_{\rm BH}-\sigma_*$ relationship for AGNs.
 This is the subject of a companion 
paper (Onken et al., this volume), which we will draw on for the 
absolute calibration of the RM-based black hole mass scale.
\end{enumerate}
The results of this program are described in detail by
Peterson et al.\ (2004), Onken et al.\ (2004), and
Kaspi et al.\ (2004) and are intended to supersede
previous compliations of AGN black hole masses by
Wandel, Peterson, \& Malkan (1999) and Kaspi et al.\ (2000).

\section{Results of the Reanalysis}
Goals of our program to improve the precision of RM-based
black hole masses have been (1) to 
improve and homogenize cross-correlation results
by employing the most up-to-date measurement and error analysis
software, and (2) to determine empirically the best measures of time
delay and line width in measurement of the central masses.
We considered 117 independent time series on 35 separate
AGNs. Our basic conclusions are:
\begin{enumerate}
\item For line-width measurement, it is important to measure
the width of the {\em variable} part of the emission line.
Our preferred method is to construct a root-mean-square
(rms) spectrum from the many spectra produced in the monitoring
campaign as the rms spectrum automatically includes only
the variable part of the emission line. Use of the mean
spectrum (or a single spectrum) is also possible, provided
that contaminants, most importantly the non-variable narrow
emission-line components, are effectively removed.
\item For time lag measurements, we considered both
the centroid and peak of the continuum/emission-line
cross-correlation function (CCF), and for the line width
measurements, we considered both FWHM and line dispersion
(i.e., second moment of the line profile). While all of
these measures are acceptable, the highest precision in virial masses
is attained by using the CCF centroid and
the line dispersion to form the virial product
$c\tau\Delta V^2/G$. 
\end{enumerate}
Using these measures, we find that the precision of RM-based black hole
mass measurements is typically about 30\%.

The masses of the central black holes are given by
\begin{equation}
\label{eq:virial2}
M_{\rm BH} = \frac{f c\tau \Delta V^2}{G},
\end{equation}
where $f$ is a factor of order unity that depends on the
structure, kinematics, and aspect of the BLR. As described
by Onken et al. in this volume, the scaling 
factor $f$ can be empirically determined by assuming that
AGNs and quiescent galaxies follow the same 
$M_{\rm BH}-\sigma_*$ relationship.
This assumption leads to  $\langle f \rangle = 5.5$.
The scatter around the correctly
scaled AGN $M_{\rm BH}-\sigma_*$ relationship is
about a factor of about 2.8, which thus
represents the typical accuracy of
RM-based mass measurements. It is important to
keep in mind that this level of accuracy is statistical
in nature and individual black hole masses may be
less accurate. 
While the limited scatter in the AGN $M_{\rm BH}-\sigma_*$ 
relationship is reassuring, additional tests 
of the efficacy of black hole mass measurement by
RM remain highly desirable.

\section{The AGN Mass--Luminosity Relationship}
Our improved black hole masses lead to the
AGN mass--luminosity relationship shown in Fig.\ 1.
We have also estimated
the bolometric luminosity in the same fashion as
Kaspi et al.\ (2000), i.e., $L_{\rm bol} \approx 
9 \lambda L_{\lambda}(5100\,\mbox{\rm \AA})$, and this scale is shown on
the upper axis of Fig.\ 1. The diagonal lines show
the Eddington limit, and 10\% and 1\% its value.
None of the sources in Fig.\ 1
exceed the Eddington limit, though of course we caution
that the bolometric correction we used is
nominal and may not apply equally well to all AGNs.
Furthermore, the optical luminosities used here have not
been corrected for the contribution of starlight from
the host galaxies --- this can be a significant factor,
especially in the lower-luminosity sources.  
Similary, no correction for internal extinction has 
been attempted.

\begin{figure}
 \includegraphics[width=14cm]{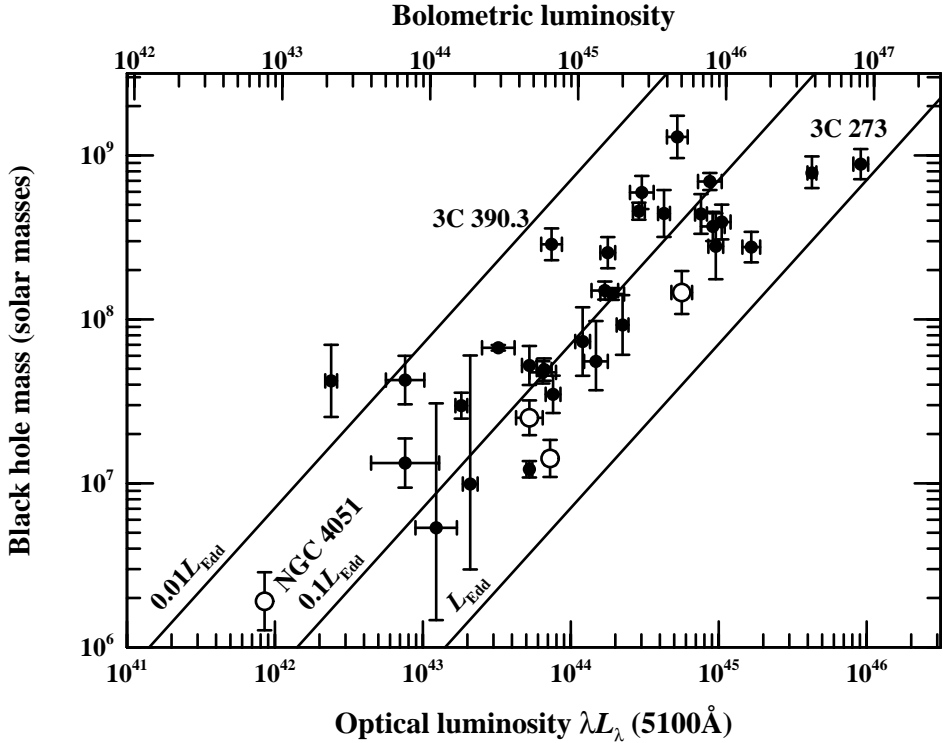}
  \caption{Black hole mass vs.\ luminosity for 35 reverberation
mapped AGNs. The luminosity scale on the lower x-axis is
$\log \lambda L_{\lambda}$ in units of ergs s$^{-1}$, and
the upper x-axis shows the bolometric luminosity. The diagonal
lines show the Eddington limit $L_{\rm Edd}$,
$0.1 L_{\rm Edd}$, and $0.01L_{\rm Edd}$. The open circles
represent NLS1s.}
\label{fig:1}
\end{figure}

Given the small formal error bars for
most of the sources, much of the scatter
in Fig.\ 1 appears to be real. Interestingly, the
scatter correlates with other AGN properties;
the narrow-line Seyfert 1 galaxies (NLS1s)
are shown as open circles, and all of them except
NGC~4051 lie on the lower edge of the mass--luminosity
envelope, along with 
several PG quasars with I Zw 1-type spectral
characteristics.
Conversely, the one reverberation-mapped object
with strongly double-peaked Balmer line 
profiles, 3C 390.3, lies along the upper
edge of the envelope. The locations of these
extreme objects on this diagram suggest that at least
some of the dispersion of the data points
correlates with Eigenvector 1, consistent with the
suggestion originated by Boroson \& Green (1992) and
reaffirmed by numerous later investigators that 
Eigenvector 1 appears to be driven by Eddington ratio
$\dot{M}/\dot{M}_{\rm Edd}$.
However, the {\em physical} origin of the scatter 
observed in Fig.\ 1 could
be attributable either to differences in Eddington ratio
or to inclination effects, or most likely some combination
of both of these effects. Decreasing
inclination (i.e., from edge-on to face-on)
will translate points to the right as the 
apparent luminosity increases on account of decreased limb darkening
and downward as the line-of-sight projection of the
rotational velocities appear to decrease.
Increasing the Eddington ratio  will
translate points in the same sense.

The best-fit slope to the relationship
$M \propto L^{\alpha}$ yields $\alpha = 0.787 \pm 0.099$.
However, there is no reason to believe that there are
no selection effects operating. Indeed, the extreme
objects 3C~273 and NGC~4051 certainly tend to make
the relationship flatter than it would be otherwise.
Interestingly, the lower edge of the envelope seems
to parallel the lines of constant Eddingtion ratio rather well,
suggesting that the intrinsic mass--luminosity
slope may not differ signficantly from unity.

\begin{figure}
 \includegraphics[width=13cm]{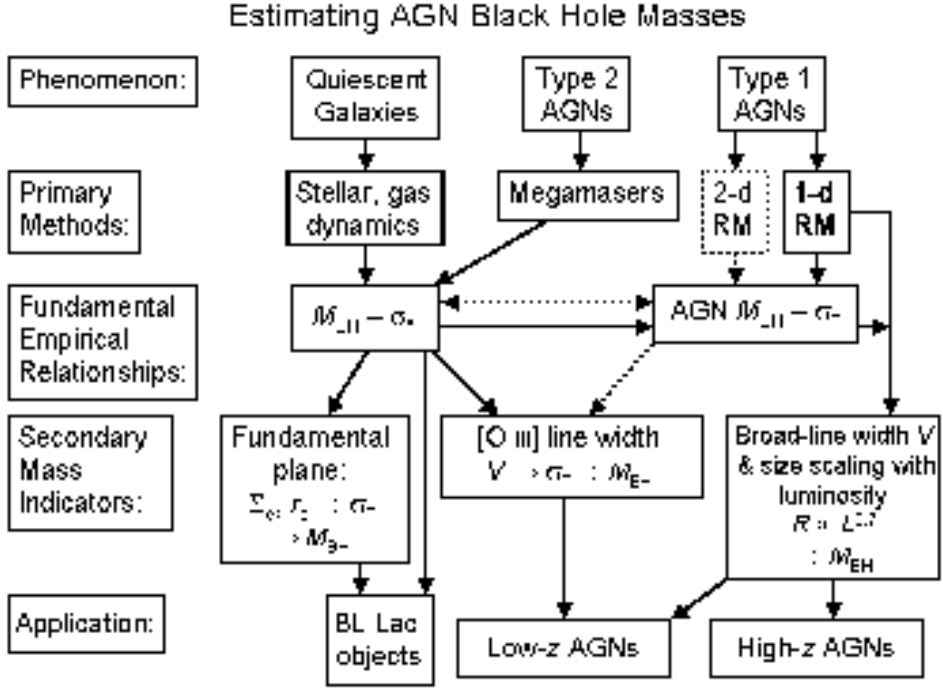}
  \caption{A summary of black hole mass measurement techniques
and how these are tied to secondary methods. The primary,
or direct, measurement methods are stellar and gas kinematics
in quiescent galaxies, megamaser motions in Type 2 AGNs, and
RM for Type 1 AGNs. ``One-dimensional (1-d) RM'' 
refers to measuring only the mean time delay for a particular
line. ``Two-dimensional (20-d) RM'' 
yields a velocity--time delay map that 
does not require the quiescent galaxy
$M_{\rm BH}-\sigma_*$ relationship to establish a zero point.
Secondary methods rely on correlations between more easily observed
properties for extension to large numbers of various
types of objects, as shown in the bottom line.}\label{fig:2}
\end{figure}

\section{Some Remarks}
\subsection{On Black Hole Masses}
In Figure 2, we summarize the methods used in measuring the masses of
black holes in galactic nuclei in an attempt to clarify the role of
RM. ``One-dimensional (1-d) RM'' 
refers to measurement of only mean time lags for
emission lines, as we have described here. Absolute
calibration of this method relies on the assumption that the
$M_{\rm BH}-\sigma_*$ relationship is the same for both active
and quiescent galaxies. ``Two-dimensional (2-d) RM,'' on the other hand,
yields a unique velocity--time delay map that reveals
the geometry and kinematics of the BLR, thus leading to 
an independent high-accuracy measurement of the black hole mass.
Unfortunately, on account of the demanding data requirements
(e.g., Horne et al.\ 2004; Peterson \& Horne 2004),
this has not yet been accomplished. The primary methods of
black hole mass measurement lead to empirical relationships 
among properties that can be used as relatively easy-to-observe
surrogates for quantities that are difficult to measure. In
the case of quiescent galaxies, the value of the stellar surface brightness 
$\Sigma_{\rm e}$ at the effective radius $r_{\rm e}$ can be used to 
infer $\sigma_*$ through the fundamental plane relationship,
and then the black hole mass can be inferred through the 
$M_{\rm BH}-\sigma_*$ relationship. In the case of active galaxies,
the narrow [O\,{\sc iii}]\,$\lambda5007$ line
width can be used as a surrogate for $\sigma_*$ (the author confesses
to deep reservations about this assumption, though practitioners
clearly note the potential problems). Another widely employed
method makes use of the correlation between BLR radius and
AGN luminosity (e.g., Kaspi et al.\ 2000). This method is especially
attractive since a single spectrum of only modest spectral resolution and 
quality can yield a luminosity (from which one infers a BLR radius)
and a line width, which can be combined for a mass estimate.
This method is already widely employed in estimating black hole
masses in large samples of AGNs, though it is clear that more work must
be done on the basic calibrations to understand 
the uncertainties more quantitatively.

\subsection{On Future Work}
Based on the existing database, we are continuing investigation of
a number of issues, notably:
\begin{enumerate}
\item Refinement of the BLR radius--luminosity relationship
(cf.\ Kaspi et al.\ 2000).
A key element here is that the luminosity measures are affected
by the starlight contribution from the host galaxy, and this is
particularly important at the low-luminosity end. We are
currently obtaining {\em Hubble Space Telescope (HST)} images of 
the lower-luminosity reverberation-mapped AGNs in order to
model the nuclear surface brightness distributions of the
starlight and thus correct the luminosities used in both
the radius--luminosity and mass--luminosity relationships.
\item Direct comparison of RM and stellar dynamical mass measurements.
Of all the reverberation-mapped AGNs, only NGC~3227 and
NGC~4151 might be expected to
have a black hole radius of influence that is
spatially resolvable with {\em HST}. These two AGNs
are thus good candidates for measuring their black hole
masses by stellar dynamical methods. A
serious problem is that the AGN nuclear light swamps the stellar
light, making the stellar absorption features difficult to detect.
We have attempted to mitigate this problem by obtaining
{\em HST} STIS spectra of the Ca\,{\sc ii} triplet in the
nucleus of NGC~4151 while the active nucleus 
was in a faint state in 2003 December.
We are now attempting to model the centeral stellar dynamics from
these data, and thus effect for the first time
a direct comparison between the RM and stellar dynamical
methods of black hole mass measurement. This is proving
to be an extremely challenging program.
\item Determine the kinematics and structure of the BLR via
RM. Reverberation mapping still offers the best promise
for understanding the BLR and understanding the potential
systematic errors in RM-based black hole masses. It is 
a technique that is still underutilized, primarily because
it is resource intensive and
moderately risky. However, the fact remains that we do
not have {\em one} reliable velocity-delay map for
{\em any} emission line in {\em any} AGN. Until we
succeed in acquiring one, we will not be able to assess
the full potential and practical merit of the technique.
\end{enumerate}

\begin{acknowledgments}
We are grateful for support of this work through NSF grant
AST-0205964 to The Ohio State University.
\end{acknowledgments}

\end{document}